\documentclass[twoside]{IEEEtran}
\usepackage[dvips]{graphicx} 
\usepackage{txfonts} 
\usepackage{cite} 
\begin{document} 
\title{Analysis of characteristics of Al MKID resonators 
\author{Takashi Noguchi, Agnes Dominjon, and Yutaro Sekimoto 
\thanks{Manuscript received September 19, 2017. }
\thanks{T. Noguchi, A. Dominjon, and Y. Sekimoto are with the Advanced Technology Center, National Astronomical Observatory, 2-1-1 Osawa, Mitaka, Tokyo, 181-8588 Japan (email: \{Takashi.Noguchi,dominjag,sekimoto.yutaro\}@nao.ac.jp).} 
\thanks{T. Noguchi is also with Department of Astronomical Science, Graduate University for Advanced Studies (SOKENDAI). } 
}}
\maketitle
\markboth{EUCAS2017 - 3EP1-04} {Noguchi \MakeLowercase{\textit{et al.}}: Analysis of characteristics of Al MKID resonators} 
\begin{abstract} 
We have prepared two kinds of Al resonators using thin films made by evaporation or sputtering and studied temperature behavior of their quality factors and resonance frequencies with the help of the extended Mattis-Bardeen (M-B) equations and its approximated analytical expressions.
We have found that temperature behavior of both the internal quality factor \boldmath{$Q_i$} and resonance frequency shift \boldmath{$\delta f_r/f_r$} measured in evaporated Al thin-film resonators are well agreed with those calculated by the analytical expressions of the complex conductivity given by the M-B theory. 
In the Al thin film resonators made by sputtering, \boldmath{$Q_i$} and  \boldmath{$\delta f_r/f_r$}  show a peak and a bump near 0.17 K in the respective temperature dependence. It is found that \boldmath{$1/Q_i$} is well approximated by \boldmath{$a - b\ln T$} below 0.15 K, where \boldmath{$T$} is the temperature.
This type of temperature dependence strongly indicates the existence of the magnetic impurity scattering of residual quasiparticles due to the Kondo effect. 
It is shown that temperature dependence of $Q_i$ and  $\delta f_r/f_r$ observed in the sputtered Al resonators can be well fitted by the theory taking the Kondo effect and kinetic inductance of the residual quasiparticles into account. 
\end{abstract}


\maketitle 
\IEEEdisplaynontitleabstractindextext 
\IEEEpeerreviewmaketitle 
\section{Introduction} 
\IEEEPARstart The quality factor of a superconducting resonator is inversely proportional to the residual quasiparticle number in the superconductor. Since the residual quasiparticle number in the superconductor is extremely reduced with decreasing temperature, it is theoretically predicted that a superconducting resonator at very low temperature with a very high quality factor $> 10^8$ can be obtained. 
However, it has been experimentally found that the quality factor of a superconducting resonator shows a saturation  with decreasing temperature at low temperatures and is strongly deviated from that predicted by the Mattis-Bardeen (M-B) theory \cite{M-B}. 
It has been widely believed that two-level states (TLS) in the dielectric are most likely responsible for the saturation of the quality factor and the deviation from the prediction of the M-B theory for a thin film superconducting resonator on a dielectric substrate \cite{Mazin,Martinis}.
Recently, authors have shown by the numerical calculations that the anomalous behavior of the quality factor of the superconducting resonator at low temperature is well predicted by the extended M-B equations in which the gap broadening effect is taken into account \cite{noguchi2016}. 

We have been experimentally studying the influence of quasiparticles in the electronic states broadened into the superconducting gap to the response of a superconducting resonator that is a key element of the microwave kinetic inductance detector (MKID).
We have prepared two kinds of Al resonators using thin films made by evaporation or sputtering and studied temperature behavior of their quality factors and resonance frequencies with the help of the extended M-B equations and its analytical expressions in order to fit to the experimental data.
We have found that temperature behavior of both the internal quality factor, $Q_i$, and resonance frequency shift, $\delta f_r/f_r$, measured in evaporated Al thin-film resonators are well agree with those calculated by the analytical expressions of the complex conductivity given by the extended M-B theory. 
In the Al thin film resonators made by sputtering, $Q_i$ and  $\delta f_r/f_r$  show a peak and a bump in the their temperature dependence near 0.17 K, respectively. 
It is found that the temperature dependence of $1/Q_i$ below 0.15 K is well approximated by $a - b\ln T$, where $T$ is the temperature. 
This type of temperature dependence strongly indicates the existence of scattering of residual quasiparticles due to Kondo effect \cite{KondoEffect}. 
In this paper, it will be shown that temperature behavior of $Q_i$ and  $\delta f_r/f_r$ observed in the sputtered Al resonators can be well predicted by the theory taking the Kondo effect and kinetic inductance of the residual quasiparticles into account.  
\section{Analysis of Al thin-film resonators}
We have prepared two types of Al thin-film superconducting resonators on high resistivity Si substrates; one was made by dc magnetron sputtering and the other was made by thermal evaporation. 
Details of the their structure and fabrication process are described in \cite{Naruse} and \cite{Agnes2016}.  
The thickness of the evaporated and sputtered Al films were 100 and 150 nm, respectively.  
The internal quality factor, $Q_i$, and resonance frequency, $f_r$, of the evaporated and sputtered Al resonators were measured in detail as function of temperature. %
\subsection{Evaporated Al thin-film resonators}
\begin{figure}[t]
\begin{center}
\includegraphics[width=0.9\linewidth]{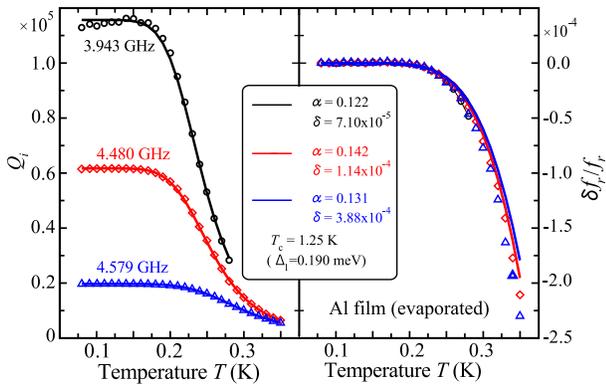}
\caption{(a) Internal quality factor $Q_i$ and (b) fractional resonance frequency change $\delta f_r/f_r$ of evaporated Al thin-film resonators as a function of temperature. Colored symbols and lines represent measured and calculated ones, respectively. Both the $Q_i$ and $\delta f_r/f_r$ calculated by the analytical expressions for the extended M-B equations agree very well with measured ones.
}
\label{fig1}
\end{center}
\vspace*{-\baselineskip}
\end{figure}

The temperature dependence of the internal quality factor $Q_i$ and fractional frequency change $\delta f_r/f_r$ of the Al thin-film resonator made by thermal evaporation are shown in Fig.~\ref{fig1}.  
The temperature dependence of the $Q_i$ and $f_r$ of the evaporated Al thin-film resonator is very similar to those reported for Al resonators \cite{Gao,Hattori}. 
In order to consider the response of the superconducting resonator to the change in
temperature, we first derived analytical approximated expressions of the complex conductivity
$\sigma_s(T)=\sigma_1(T) -i\,\sigma_2(T)$ of a superconductor given by the extended M-B theory, assuming that the gap parameter is a complex number of $\Delta=\Delta_1 +i\,\Delta_2 \ (\Delta_2 \ll \Delta_1)$.  
The real and imaginary part of the complex conductivity, $\sigma_1(T)$ and
$\sigma_2(T)$, have the following analytical approximate formulas under the condition
that $\hbar\omega \ll \Delta_1$ and $k_B T \ll \Delta_1$; 
\begin{eqnarray}
\frac{\sigma_1(T)}{ \sigma_{N}} & \simeq  &
\frac{4 \Delta_1}{\hbar \omega} 
\exp\left(-\frac{\Delta_1}{k_B T}\right)
\sinh\left(\frac{\hbar\omega}{2 k_B T}
\right)\, K_0 \left(\frac{\hbar\omega}{2 k_B T}\right)  \nonumber \\ 
& &  + \frac{\pi\Delta_2}{\hbar \omega} 
\Biggl[ 1 + \frac{2 \Delta_1}{k_B T} 
\exp\left(-\frac{\Delta_1}{k_B T}\right)  \nonumber \\
& & \hspace*{2cm} \times 
\exp\left(-\frac{\hbar\omega}{2k_B T}\right)\, 
I_0 \left(\frac{\hbar\omega}{2 k_B T}\right)\Biggr] \label{s1}  \\ 
\frac{\sigma_2(T)}{ \sigma_{N} } & \simeq & \frac{\pi  \Delta_1}{\hbar \omega} 
\Biggl[ 1- 2 \exp\left(-\frac{\Delta_1}{k_B T}\right)  \nonumber \\
& & \hspace*{2cm} \times 
\exp\left(-\frac{\hbar\omega}{2k_B T}\right)\, 
I_0 \left(\frac{\hbar\omega}{2 k_B T}\right)\Biggr] , \label{s2}
\end{eqnarray}
where $\sigma_N$ is a normal conductivity just above $T_c$ and $I_0(x)$ and $K_0(x)$ are the 0-th
order modified Bessel function of the first and second kind with the argument $x$, respectively.  Note here that
$\Delta_1$ and $\Delta_2$ are assumed to be constant at temperatures well below $T_c$. Then the $Q_i$ and $\delta f_r/f_r$ are calculated by
\begin{eqnarray}
Q_i(T)  & = & \frac{1}{\alpha}\frac{\sigma_2(T)}{\sigma_1(T)} \label{Qi} \\
\frac{\delta f_r}{f_r} & \equiv & \frac{f_r(T) - f_r(0)}{f_r(0)} \ \  = \ \ -\frac{1}{2}\alpha\frac{\delta \sigma_2(T)}{\delta \sigma_2(0)} , \label{dfr0} 
\end{eqnarray}
where
$\alpha$ is a ratio of the kinetic inductance $L_K$ to the total inductance of the resonator $L$.
The solid lines in Fig.~\ref{fig1} are $Q_i$ and  $\delta f_r / f_r$ calculated by using eqs. (\ref{s1})-(\ref{dfr0}). In the calculations, fitting parameters $\Delta$ and $\alpha$ are determined  so as to give the best fits to the measured data.
It is found that the temperature dependence of the $Q_i$ and $f_r$ of the evaporated Al thin-film resonator are well fitted by the analytic expressions of the extended M-B equations given by eqs.~(\ref{s1}) and (\ref{s2}).

\subsection{Sputtered Al thin-film resonators}
Examples of the measured internal quality factor $Q_i$ and fractional frequency change $\delta f_r/f_r$ of sputtered Al thin-film resonators are shown in upper and lower panels in Fig.~\ref{fig1}, respectively.  
It is found that $ Q_i $ of the sputtered Al thin-film resonator sharply increases with decreasing temperature, reaches a peak at about 0.17 K, and decreases as the temperature further decreases. On the other hand, it is found that as temperature decreases, $ \delta f_r/f_r$ rapidly increases, sign changes from negative to positive at about 0.22 K, and then gradually decreases after showing peak at around 0.17 K.
Since the temperature dependence of both $Q_i$ and $\delta f_r/f_r$ of the sputtered Al thin-film resonators
are quite different from those of evaporated resonators and 
it seems to be difficult to fully explain the temperature dependence of the $Q_i$ and $\delta f_r/f_r$ observed in sputtered Al thin-film resonator only by the extended M-B theory so that further theoretical and analytical consideration were made as described below.  
%
%
\subsubsection{Quality factor}
Inverse of the measured internal quality factors $1/Q_i$ of an sputtered Al resonator are plotted as a function of temperature below 0.2 K in Fig.~\ref{fig3}.  
It is clearly shown that $1 / Q_i$ below 0.15 K is well approximated by the formula $a - b \ln T = -b \ln(T/T_K)$ as shown in Fig.~\ref{fig3}, where $a$, $b$ and $T_K$ are fitting parameters and $T_K$ is called Kondo temperature. 
Since  $1/Q_i $ is approximately given by $\alpha (R_{res} / \omega_r L_K)$ and $\omega_r L_K$ is a constant at this temperature, this result indicates that the residual resistance $R_{res}$ below  0.15 K shows Kondo effect-like temperature dependence, and also strongly suggests the existence of magnetic impurity scattering of the residual quasiparticles due to the Kondo effect in the superconducting resonator, where $\omega_r=2\pi f_r$.
At $T > 0.25$ K, the temperature dependence of $1 / Q_i$ is well agreed with the prediction of the extended M-B theory. 
%
\begin{figure}[t]
\begin{center}
\includegraphics[width=0.7\linewidth]{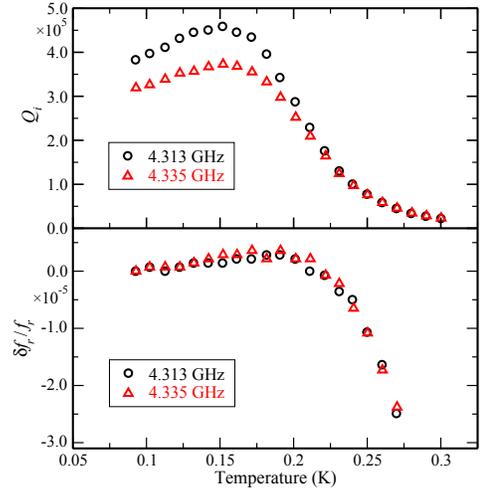}
\caption{Measured internal quality factor $Q_i$ (upper panel) and fractional frequency change $\delta f_r / f_r$ (lower panel) of sputtered Al thin-film resonators as a function of temperature. 
}
\label{fig2}
\end{center}
\vspace*{-\baselineskip}
\end{figure}
%

Taking the $\ln T$-temperature dependence of $1/Q_i$ above mentioned into consideration, the total resistance of the superconducting resonator at well below $T_c$ can be approximately written as sum of surface resistance $R_s$ and quasiparticle resistance $R_{res}$, where $R_{res}$ has $a - b \ln T$ or $-b \ln(T/T_K)$-type temperature dependence. Thus
\begin{eqnarray}
  \frac{1}{Q_i} & = & \frac{R_{s}}{\omega_r L} +
  \frac{R_{res}}{\omega_r L}\nonumber\\
  & = & \alpha \frac{\sigma_1 ( T)}{\sigma_2 ( T)}
 - b \ln \left( \frac{T}{T_K} \right)  , \label{kondo_eq}
\end{eqnarray}
where $T_K$ is a Kondo temperature of the quasiparticle system in the resonator. 
%
\begin{figure}[t]
\begin{center}
\includegraphics[width=0.8\linewidth]{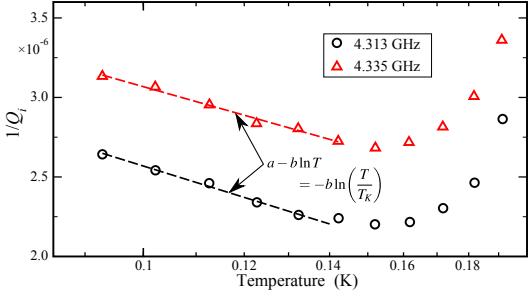}
\caption{Inverse quality factor $1/Q_i$ as a function of temperature below 0.2 K. Kondo effect-like temperature behavior in $1/Q_i$ are clearly observed.
}
\label{fig3}
\end{center}
\vspace*{-\baselineskip}　
\end{figure}
Using eq.~(\ref{kondo_eq}), we calculated $1 / Q_i$ and fitted to the experimental data using $\alpha$, $b$ and $T_K$ as fitting parameters.  
In the case of a sputtered aluminum resonator, from the fitting of $1/Q_i$ measured at a temperature less than 0.15 K, the slope $b$ and the Kondo temperature $T_K$ can be determined independently.
The fitted curves to the data are plotted by solid lines in Fig.~\ref{fig4}. It is demonstrated 
that very nice agreement between $Q_i$ calculated by eq.~(\ref{kondo_eq}) and measured ones is obtained. 

\subsubsection{Resonance frequency variation}
It was indicated in the previous sections that the quasiparticles of the sputtered Al resonator at $T<0.15\ $K are in the Kondo state. The  Kondo state is characterized by the  Kondo correlation length \cite{Campagnano} given by 
\begin{equation}
\xi_K = \frac{\hbar v_F}{k_B T_K} , \label{Kondo_length}
\end{equation}
where $v_F$ is Fermi velocity. Thus the lifetime of quasiparticles $\tau_K$ in the Kondo state is given by
\begin{equation}
\tau_K = \frac{\xi_K}{v_F} = \frac{\hbar}{k_B T_K} . \label{Kondo_lifetime}
\end{equation}
From the fitting of $1/Q_i$ in the previous sections, we obtained $T_K \simeq 0.65\ $ K, which gives the lifetime of $\tau_K \approx $12 ps. 

In the Drude model the ac resistivity of the quasiparticle system with a mean free time $\tau$ at a frequency $\omega$ is given as 
\begin{equation}
  \rho = \rho_0 ( 1 + i \omega \tau) , \label{drude_resistance}
\end{equation}
where $\rho_0$ is a dc resistivity and $\rho_0 \tau$ is called kinetic inductance of the qusiparticles. 
Assuming that the quasiparticles are in the Kondo state and that microwave readout signal with a frequency $\omega_r$ is coupled to the resonator, $\tau$ in eq.~(\ref{drude_resistance}) is equal to $\tau_K$ and $\omega_r\tau_K \approx 0.25\ $ for $\omega_r/2\pi = $4 GHz, which is not negligibly small that we have to take into consideration the imaginary part, or kinetic inductance component in eq.~(\ref{drude_resistance}) . Thus, 
we have taken into account the contribution of the kinetic inductance component of the resistivity to the total surface
impedance. The total surface impedance $Z_s$ of a superconductor is given as
\begin{eqnarray}
  Z_s = R_s ( 1 + i \omega \tau) + i X_s = R_s + i (X_s +x_s)
\end{eqnarray}
where $R_s$ and $X_s$ are the surface resistance and reactance of the superconductor given by the M-B theory, respectively and $x_s$ is a reactance of quasiparticles defined as $x_s =\omega_r \tau R_s$. 

Now we consider a superconducting resonator such as that made of Al and modify
the calculation of its electrical properties taking the contribution of
quasiparticle kinetic inductance into account. Especially 
the resonance frequency shift must be calculated by replacing  $X_s  \rightarrow  X_s + x_s $, and we obtain 
\begin{equation}
  \frac{\delta X_s}{X_s} \rightarrow \frac{\delta X_s + \delta x_s}{X_s + x_s}  \simeq   \frac{\delta X_s}{X_s} + \omega_r \tau\, \delta\frac{1}{Q_s} \label{dXs/Xs} ,
\end{equation}
where $\delta\frac{1}{Q_s} =1/Q_s(T) - 1/Q_s(0)$. 
Since $Q_i \simeq Q_i(0) \simeq 10^6$ for the Al resonator, if $\omega_r \tau \,{\ge}\, 0.25$ for $T_K \simeq 0.65$ K,  
the second term in eq. (\ref{dXs/Xs}) is almost comparable in magnitude for the first term of $\delta X_s/X_s \sim 10^{- 7}$ at the temperature below 0.15 K. 
Then the fractional frequency change, $\delta f_r / f_r$, of the superconducting resonator is given as 
\begin{equation}
 \frac{\delta f_r}{f_r}  
 = - \frac{1}{2} \alpha \frac{\delta X_s + \delta x_s}{X_s} 
 = \frac{1}{2} \left\{ \alpha  \frac{\delta \sigma_2}{\sigma_2}
 - \omega_r \tau\, \delta\frac{1}{Q_i} \right\} . \label{dfr}
\end{equation}
%
\begin{figure}[t]
\begin{center}
\includegraphics[width=0.8\linewidth]{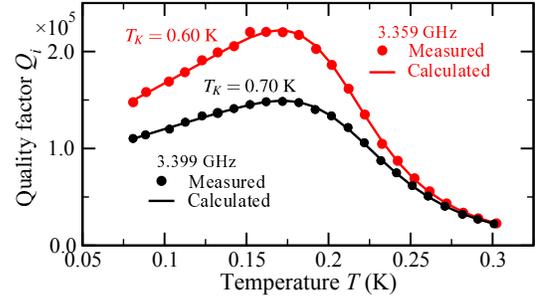}
\caption{Measured and calculated $Q_i$ as a function of temperature. Symbols and lines represent measured data and fitted curves, respectively. Kondo temperatures $T_K$  obtained by the fitting are also shown.
}
\label{fig4}
\end{center}
\vspace*{-\baselineskip}
\end{figure}

\begin{figure}[t]
\begin{center}
\includegraphics[width=0.7\linewidth]{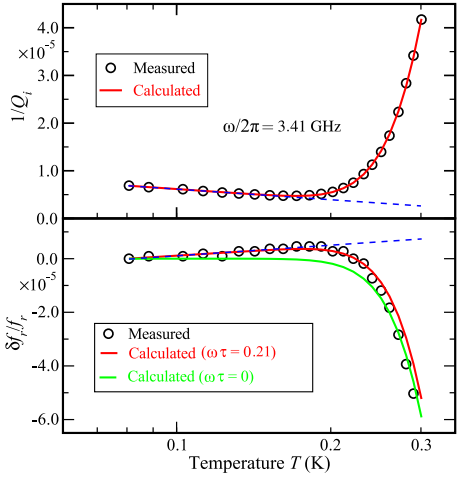}
\caption{Inverse of quality factor $1/Q_i$ (upper panel) and fractional resonance frequency change $\delta f_r/f_r$ (lower panel) of an sputtered Al resonator as a function of temperature in logarithmic scale. Open circles are measured ones and solid lines in red plot calculated ones using eqs. (\ref{kondo_eq}) and (\ref{dfr}) to give the best fits to the measured data. Solid line in green represents a plot for $\tau =0$ in eq. (\ref{kondo_eq}) for reference. Broken lines in blue shows $\ln T$ temperature dependence.}
\label{fig5} 
\end{center}
\vspace*{-\baselineskip}
\end{figure}

\begin{figure}[t]
\begin{center}
\includegraphics[width=0.7\linewidth]{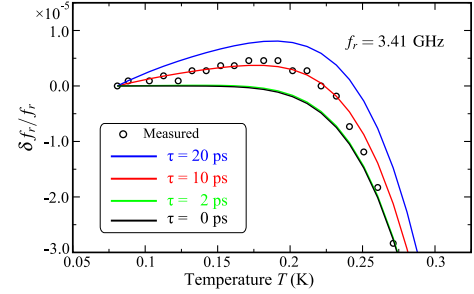} 
\caption{Fractional resonance frequency change $\delta f_r/f_r$ of a sputtered Al  resonator as a function of temperature in logarithmic scale. Open circles are measured ones and solid lines show the calculated ones using eqs. (\ref{kondo_eq}) and (\ref{dfr}) for several lifetime $\tau$.  }
\label{fig6}
\end{center}
\vspace*{-\baselineskip}
\end{figure}

The inverse of quality factor $1/Q_i$ and fractional frequency change $\delta f_r / f_r$ calculated by
using eqs. (\ref{kondo_eq})  and (\ref{dfr}) for fitting and the measured data are shown in Fig.\ref{fig5} (a) and (b) , respectively. As was also shown in Fig. \ref{fig4} for the fitting of $Q_i$, a very nice agreement between the measured and calculated $1/Q_i$ is obtained.  
It is also found that a good agreement between the calculated and measured $\delta f_r / f_r$ is achieved, when 
$\omega_r \tau \simeq 0.21$, which corresponds to $\tau \simeq10\ $ ps for $\omega_r/2\pi=3.41$ GHz and is quite consistent with $\tau_K\sim$12 ps expected from eq.~(\ref{Kondo_lifetime}) for $T_K=0.65\ $ K. This result indicates that the $\ln T$ behavior and the bump below 0.25 K in $\delta f_r/f_r$ is due to the contribution of the kinetic inductance of quasiparticles.  In Fig.~\ref{fig6}, fractional resonance frequency change $\delta f_r/f_r$ as a function of temperature for several quasiparticle lifetime are shown together with measured data. As already shown in Fig.~\ref{fig5}, the solid line in red calculated for $\tau =10\ $ agrees well with the measured data. By looking at the fitted data carefully, however, a little discrepancy between the measured and calculated $\delta f_r / f_r$ is found at temperature above $\sim$0.25 K.
Since the Kondo scattering is superseded by the other scattering mechanism, such as the phonon scattering, above $\sim$0.25 K, the lifetime of the quasiparticles might be shorter than that in the Kondo state below $\sim$0.25 K. As a result the measured data becomes in agreement with calculated ones for $\tau \rightarrow 0$ above $\sim$0.25 K as shown in Fig.~\ref{fig6}. It is noted here that the height of the bump of $\delta f_r/f_r$ below $\sim$0.25 K increases as the quasiparticle lifetime increases. This means that the contribution of the quasiparticle kinetic inductance to the resonace frequency increases as quasiparticle lifetime increases.

\section{Discussion}
Since the Kondo effect is a scattering mechanism of conduction electrons due to magnetic impurities, the difference of the quality factor as a function of temperature between the evaporated and sputtered Al film resonators might be attributed to the difference of densities of magnetic impurities in those films. 
Although we think that those magnetic impurities mainly come from source material, some of them especially in the sputtered film are contaminated from the vacuum system during deposition. 
Since it is necessary to know the concentration of the magnetic impurities in those films, we plan to make detailed impurity analysis in near future.

A very similar temperature dependence of both $Q_i(T)$ and $\delta f_r(T)/f_r$ in an Al film resonator to those observed in our sputtered Al film resonator has been reported and analyzed by using the theory in which the RF loss is  postulated to arise from coupling to TLS defects in the dielectric \cite{McCarrick}. Although it has been shown that the measured $\delta f_r(T)/f_r$ is in good agreement with the expectation from the TLS theory, the agreement between $Q_i(T)$ predicted by the TLS theory and measured one seems to be rather poor.
Unlike the conventional TLS theory, the analysis of the Al resonators presented here are only based on the theory on the electronic properties of a superconductor, especially on the assumption of the existence of residual quasiparticles at temperature well below $T_c$. It was successfully demonstrated that the prediction of temperature behavior of both quality factor and resonance frequency shift by this theory agree very well with the experimental data. To get a superconducting resonator with a very high quality factor, more detailed investigation and further extension of the present theory are needed to make clear the loss mechanism in the superconducting resonator. 
  
\section{Summary}
It is demonstrated that quality factors and resonance frequency shifts  as a function of temperature experimentally obtained in evaporated Al thin film resonators are good agreements with those predicted by the extended M-B theory. 
It is found that inverse quality factors of sputtered Al thin film resonators show $a-b \ln T$ temperature dependence at the temperature below 0.15 K, which indicates the existence of the Kondo effect in the residual quasiparticle system.  
Assuming that the residual quasiparticles are in the Kondo state, due to its long lifetime, kinetic inductance of quasiparticles becomes so large that it seriously contributes to the  resonance frequency shift at the temperature where the Kondo effect-like behavior is observed. 
It is shown that the quality factor and  fractional frequency changes as a function of temperature measured in the sputtered Al resonators are in very good agreement with those expected by the theory taking the Kondo effect and the kinetic inductance contribution into consideration.

\end{document}